# Comment on "The effect of the charge density of microemulsion droplets on the bending elasticity of their amphiphilic film", J. Chem. Phys. **114**, 10105 (2001)

V. Lisy

Department of Biophysics, P.J. Safarik University,
Jesenna 5, 041 54 Kosice, Slovakia

**Abstract**. It is shown that the work by Farago and Gradzielski [J. Chem. Phys. **114**, 10105 (2001)] is based on incorrect expressions for the scattering functions, contains a number of other serious defects, and should be revised.

In the recent paper by Farago and Gradzielski[1] the effect of the charge density on the bending elasticity of the amphiphilic film of microemulsion droplets has been studied. Below we show that the determination of the parameters characterizing the surface layer of the droplets from neutron scattering experiments contains serious shortcomings and should be completely revised. First of all, the interpretation of the experiments is based on incorrect expressions for the scattering functions. In Section III.C the authors derive a formula for the intermediate scattering function $I(q,t)$ that describes the quasielastic scattering of neutrons on a sphere fluctuating in the shape. The function $I(q,t)$ consists of three parts. The first term describing the scattering from a sphere in the absence of the fluctuations is $\sim P_{stat}(q) = \rho^2 [R_0^2 j_1(qR_0)/q]^2$. Thus the radius $R_0$ here is the radius of a nonfluctuating sphere. The second term $P_{stat\_corr}(q)$, that gives a time independent correction if the fluctuations are present, is then determined incorrectly. To explain this statement we note that the expansion of the radius $R(\vartheta,\varphi)$ of the deformed droplet[1], $R(\vartheta,\varphi) = R_0 \left[1 + \sum_{lm} u_{lm} Y_{lm}(\vartheta,\varphi)\right]$, is used in the calculation of $I(q,t)$. If some quantity is calculated to the first order in the fluctuation amplitudes $u_{lm}$, the $l=0$ and $l=1$ modes can be excluded from the consideration since they are of higher order in the small amplitudes $u_{lm}$.[2] For the correct determination of $I(q,t)$ to the second order in $u_{lm}$, however, all the modes should be taken into account. While the $l=1$ mode drops out during the calculation, the mode $l=0$ contributes through the coefficient $u_{00}$. This contribution is missed in Eq. (12). The amplitude $u_{00}$ is found from the condition of the conservation of the droplet volume, $V = 4\pi R_0^3/3$, hence $u_{00} = -(4\pi)^{-1/2} \sum_{l>1,m} |u_{lm}|^2$. If this is not taken into account, the volume would be

$$V\left[1 + (3/4\pi)\sum_{l>1,m}|u_{lm}|^2\right],$$

the surface area

$$A = 4\pi R_0^2 + (R_0^2/2)\sum_{l>1,m}|u_{lm}|^2(l^2 + l + 2)$$

instead of the correct

$$A = 4\pi R_0^2 + (R_0^2/2)\sum_{l>1,m}|u_{lm}|^2(l-1)(l+2),$$

*etc*. The formulae used in Ref.[1] for $\langle |u_{lm}|^2 \rangle$ have been also derived assuming the conservation of the volume.[3] Equation (14) for the scattering from a multishell structure[1] is also incorrect since again the term $\langle u_{00} \rangle$ is missed in $P_{stat\_corr}(q)$. The expression for $I(q,t)$ describing the scattering from a droplet covered with a shell of arbitrary thickness and fluctuating in the shape has been found in our earlier paper.[4] The term corresponding to $P_{stat\_corr}(q)$ is

$$P_{stat\_corr}(q) = \frac{R_0}{q} \sum_{l>1} \frac{2l+1}{4\pi} \langle u_{l0}^2 \rangle [(\rho_0 - \rho_1)R_1^2 j_1(x_1) + (\rho_2 - \rho_0)R_2^2 j_1(x_2)]$$

$$\times \{(\rho_0 - \rho_1)R_1[dj_0(x_1) - R_0 x_1 j_1(x_1)] - (\rho_2 - \rho_0)R_2[dj_0(x_2) + R_0 x_2 j_1(x_2)]\},$$

where the indices $i=0$, 1 and 2 in the scattering length densities $\rho_i$ refer to the shell, and the interior and exterior of the droplet, respectively, $x_i=qR_i$, $R_1$ is the inner, $R_2$ the outer radius of the shell, and $R_0$ is the mean of the radii. When $d<<R_0$ and for the perfect shell contrast, $\rho_1=\rho_2=\rho$,

$$P_{stat\_corr}(q) \approx -[R_0^2 d(\rho - \rho_0)]^2 J(x) \sum_{l>1} \frac{2l+1}{4\pi} \langle u_{l0}^2 \rangle,$$

with $J(x)=(x^2+2)j_0^2(x)$, while from Ref.[1] one finds $J(x)=2j_1(x)j_0(x)+(x^2-2)j_0^2(x)$. The term $P_{stat\_corr}(q)$ is absent in the work[3]. It is also either missed in previous works by Farago and coworkers, e.g.[7], or presented incorrectly, and even differently[6,8] from the above expression following from Ref.[1].

The discussed work contains other serious defects. So, from the methodical point of view the determination of the shell parameters from the fits to SANS data is flawed. The authors first neglect the fluctuation contribution and obtain an estimate for the polydispersity $p^2$ and the mean radius $R_m$. With these values the fluctuations are calculated and in the second turn the parameters are refined. In the correct approach however all the parameters $R_m$, $p^2$, $\kappa$, and $d$ should simultaneously enter each fit onto the SANS data as it was done in our analysis.[4] The method of Ref.[1] would be correct only in the case when the fluctuation contribution is negligible, as it is for large $\kappa$. However, one cannot assume it from the beginning. When the fluctuations are not neglected, the fit to the experiment could yield a different set of the parameters, e.g. with $R_m$ larger and/or $\kappa$ smaller than in the case of no fluctuations. Such a possibility was found and discussed in Ref.[4]. Since the simultaneous determination of all the parameters from SANS will lead (if the fluctuations play a role) to different $R_m$, $p^2$, and $d$, one cannot be sure that adjusting $\kappa$ from NSE, its value remains the same as determined by the discussed method.[1] As to the NSE, another shortcoming in its description is that the different viscosities of the bulk fluids are not taken into account, although it has been already shown that the account for this difference is essential.[5] Note also that in previous works by Farago and coworkers[6] a peak observed in the $q$ dependence of the effective diffusion coefficient $D_{eff}(q)$ of the droplets has been analyzed. Unfortunately, in Ref.[1] only the normalized $I(q,t)/I(q,0)$ is presented. It would be interesting to compare $D_{eff}(q)$ (which is more sensitive to the value of $\kappa$) with the experiment and thus judge the validity of the theory. We propose this because we have shown[4] that the height of the observed peak in $D_{eff}(q)$ is in a sharp disagreement with the previous theories if $\kappa \sim 1$ $kT$ or larger (notice the value 3.6 $kT$ de-termined by Farago and Gradzielski[1]). Continuing the criticism we note that the averaging over the droplet distribution in radii is done using the Schulz distribution.[1] Since the measured signal depends strongly on the droplet radius, it is sensitive also to the polydispersity. Thus it is important to use a correct distribution function. Based on the microemulsion thermodynamics, from $p^2$ the combination $2\kappa+\bar{\kappa}$ is determined. The authors[1] employ the Schulz distribution function, however the distribution that follows from the free energy of the microemulsion is $f(R_0) \propto \exp[-(1-R_0/R_m)^2/2p^2]$ with $p^2$ obeying the equation $p^2=kT/[8\pi(2\kappa+\bar{\kappa})+2kTF(\Phi)]$ used in Ref.[1]. Finally, the mean radius $R_m=\langle R_0 \rangle=\langle (R_1+R_2)/2 \rangle$ as used in the Schulz distribution[1] does not correspond to the hard sphere diameter which the authors determine as $2(R_m+d+2.5\text{Å})$ while it should be $2(R_m+d/2+2.5\text{Å})$ This again brings an error in the calculations since the difference is relatively large ($d\approx 11\text{Å}$).

In conclusion, the work by Farago and Gradzielski is based on the incorrect theory, contains a number of serious defects, and should be completely revised. The obtained value of the bending elasticity coefficient $\kappa=3.6\ kT$ (that contradicts to all other results from the literature except the previous investigations by Farago and coworkers, e.g.[6,8]) cannot be considered as reliable.